%% file: article.tex
\documentclass[12pt]{article}
\usepackage{epsfig}

\input{econfmacros.tex}
\def\Title#1{\begin{center} {\Large {\bf #1} } \end{center}}

\usepackage{slashed}
\begin{document}

\Title{Search for New Physics in Single Top Channel at the LHC}
\bigskip\bigskip


\begin{raggedright}  

{\it Muhammad Alhroob, on behalf of the ATLAS and CMS Collaborations\index{Alhroob, M.}\\
ICTP INFN Gruppo Collegato di Udine\\
Udine, ITALY}
\bigskip\bigskip
\end{raggedright}
\section{Introduction}
At the seventh International Workshop on the CKM Unitarity Triangle, a talk on the search for new physics in single top quark production at the LHC experiments was presented. Three analysis were shown: the search for single top quark production through FCNC processes performed by the ATLAS~\cite{Aad:2008zzm}, the search for $W^{'}$ boson by ATLAS using the $tb$ resonance using the invariant mass distribution of the $tb$ system  and the search for $W^{'}$ boson done by CMS~\cite{Chatrchyan:2008aa} using boosted decision tree and the invariant mass distribution of the $tb$ system. 

\section{Search for FCNC in Single Top Quark Production at ATLAS}
Flavour changing neutral currents (FCNC) are strongly suppressed in the Standard Model, especially when the top quarks are involved. The FCNC branching fractions for the top quark decaying into gauge bosons and light quarks, e.g. $t\rightarrow q$+gluon,
are predicted to be less than $10^{-13}$~\cite{Eilam:1990zc}.
However, there are extensions of the SM which predict the presence of FCNC contributions with significantly enhanced  rates.
Using the data collected in 2011 with an integrated luminosity of 2.05 ${\rm fb}^{-1}$ at $\sqrt{s}= 7$~TeV,  ATLAS studied direct single top quark production through FCNC $qg\rightarrow t\rightarrow bW$~\cite{Aad:2012gd}, where the u(c) quark interacts with a gluon to produce a single top quark which in turn decays as predicted in the SM process, $t\rightarrow W+b$.

Only the leptonic decay of the $W$ boson was considered and therefore events were selected using a single lepton ($e$ or $\mu$) trigger.
The events were required to have only one reconstructed well isolated lepton with $p_{\rm T}>25$~GeV. The missing transverse momentum, 
$\slashed E_{\rm T}$, is required to be greater than 25~GeV. To suppress the background further, the sum of $\slashed E_{\rm T}$ and $m_{{\rm T},W}$ was required to be greater than 60~GeV where $m_{{\rm T},W} = \sqrt{ 2 p_{{\rm T},\ell} \slashed E_{\rm T} (1- \cos(\Delta\phi(\ell, \slashed E_{\rm T}))}$. 
Given that the top quark decays to a $W$ and a $b$ quark, events were selected if they contained exactly  one 
reconstructed jet with $p_{\rm T}>25$~GeV, which was identified as a jet originated from a $b$ quark. 

Given the large amount of the background with very large uncertainties, and the small number of expected signal events, 
a multivariate analysis technique was needed to isolate the signal from background.
A neural network was used that combines a three-layer feed forward neural network with complex robust preprocessing. The neural network was trained using 11 input variables, where the most discriminating were the $p_{\rm T}$ of the reconstructed $W$ boson, the distance in $\eta$-$\phi$, $\Delta R(\rm {b-jet,\ell})$, and the lepton electric charge. The hidden layer consisted of 
13 nodes and one output node with a continuous output in the interval [$-1, 1$].

To calculate the signal cross section, a Bayesian statistical method was used, with a binned likelihood of the full neural network output distribution.  
The signal prior was chosen to be flat. Rate normalisation effects and shape distortions of the template distributions were included with a direct 
sampling approach using Gaussian priors with the mean equal to 0 and width equal to 1. No evidence for  FCNC single top quark events was found and 
an upper limit on the cross section at 95\% C.L. was calculated to be 3.9~pb.

\begin{figure}
\begin{center}
\epsfig{file=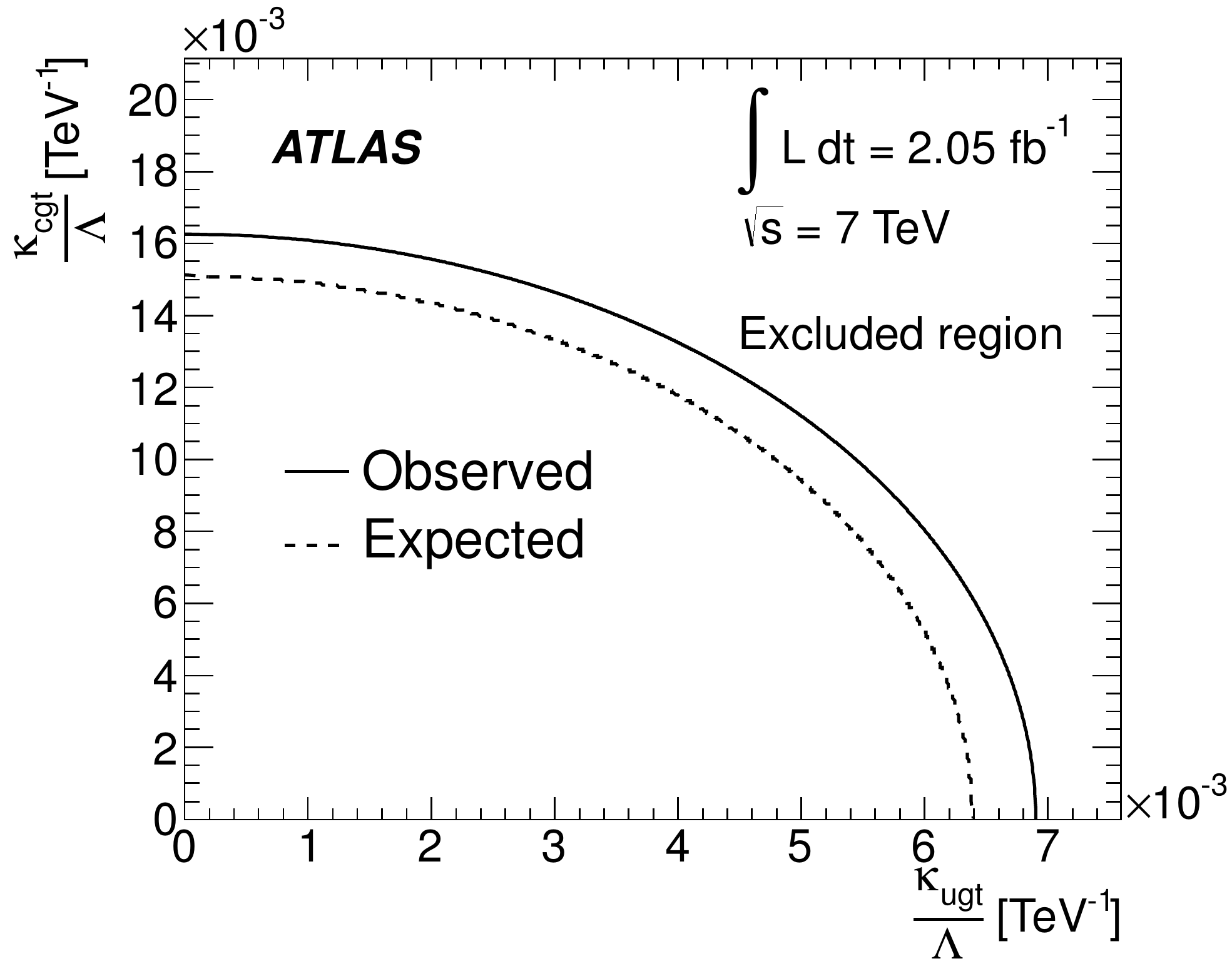,height=2.3in}
\epsfig{file=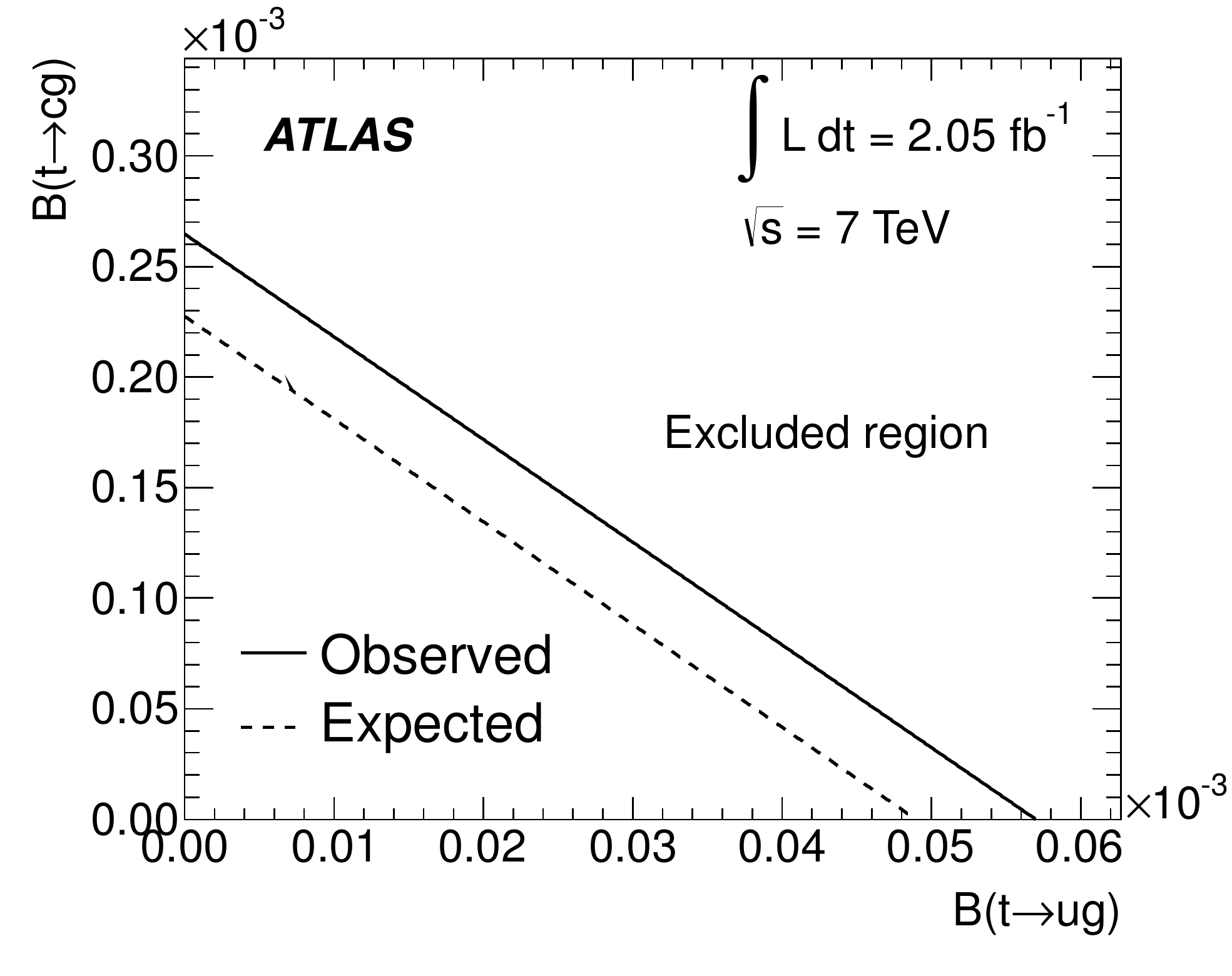,height=2.3in}
a)\hspace*{0.5\textwidth} b)
\caption{\label{fig:finalpdf}
a) The upper limits at 95\% C.L. on the coupling strengths $\frac{\kappa_{ugt}}{\Lambda}$ and $\frac{\kappa_{cgt}}{\Lambda}$. b) The upper limits at 95\% C.L. on branching fractions of $t \rightarrow ug$ and $t \rightarrow cg$~\cite{Aad:2012gd}.}
\end{center}
\end{figure}

The cross section upper limit was converted to an upper limit on the coupling strengths,$\frac{\kappa_{ugt}}{\Lambda}$ and $\frac{\kappa_{cgt}}{\Lambda}$, using a NLO calculation following model independent approach~\cite{Gao:2011fx}. Assuming $\frac{\kappa_{cgt}}{\Lambda} = 0$ the $\frac{\kappa_{ugt}}{\Lambda} < 6.9\times10^{-3}$ TeV$^{-1}$, assuming $\frac{\kappa_{ugt}}{\Lambda} = 0$ the $\frac{\kappa_{cgt}}{\Lambda} < 1.6\times 10^{-2}$~TeV$^{-1}$. Similarly, the upper limits on the coupling strengths were converted to limits on branching fractions using NLO calculation~\cite{Zhang:2008yn}, resulting in Br$(t \rightarrow ug) < 5.7\times 10^{-5}$ assuming Br$(t \rightarrow cg)=0$, and Br$(t \rightarrow cg) < 2.7\times 10^{-4}$ assuming Br$(t \rightarrow ug)=0$. Figure~\ref{fig:finalpdf}a and Figure~\ref{fig:finalpdf}b show  the distributions of  the upper limits for all possible couplings and branching fractions at 95\% C.L., respectively. 

\section{Search for $W^{'}$ at ATLAS and CMS}

Various extensions to the SM predict extra massive charged bosons, called $W^{'}$. In contrast to the SM $W$ boson, the $W^{'}$ boson can couple to fermions with pure right, left handed couplings or a mixture of the two according to the model.  

ATLAS searched for the $W^{'}$ boson using 1.04~${\rm fb}^{-1}$ of data collected in 2011 at $\sqrt{s}=7$~TeV~\cite{Aad:2012ej}. Many models predict the $W^{'}$ boson to be coupled more strongly to the third generation of quarks than the first and the second generations, thus the search for the $W^{'}$ boson was performed by looking at the $tb$ resonance. The analysis was performed under the assumption that the $W^{'}$ boson couples only to the right handed fermions with the SM model like couplings.

The top quark was assumed to decay 100\% to a $W$ boson and a $b$ quark, where the leptonic decay of $W$ was considered.  $W^{'}$ boson events were selected by applying the same 
cuts as used to select the FCNC candidate events except that events were required to have only two high $p_{\rm T}$ jets with at least one of them originated from a $b$ quark.
Two channels were identified: events with exactly one $b$ quark jet where the dominant background was $W$+jets, and events with two $b$ quark jets where the dominant background was $t\bar{t}$.

\begin{figure}
\begin{center}
\epsfig{file=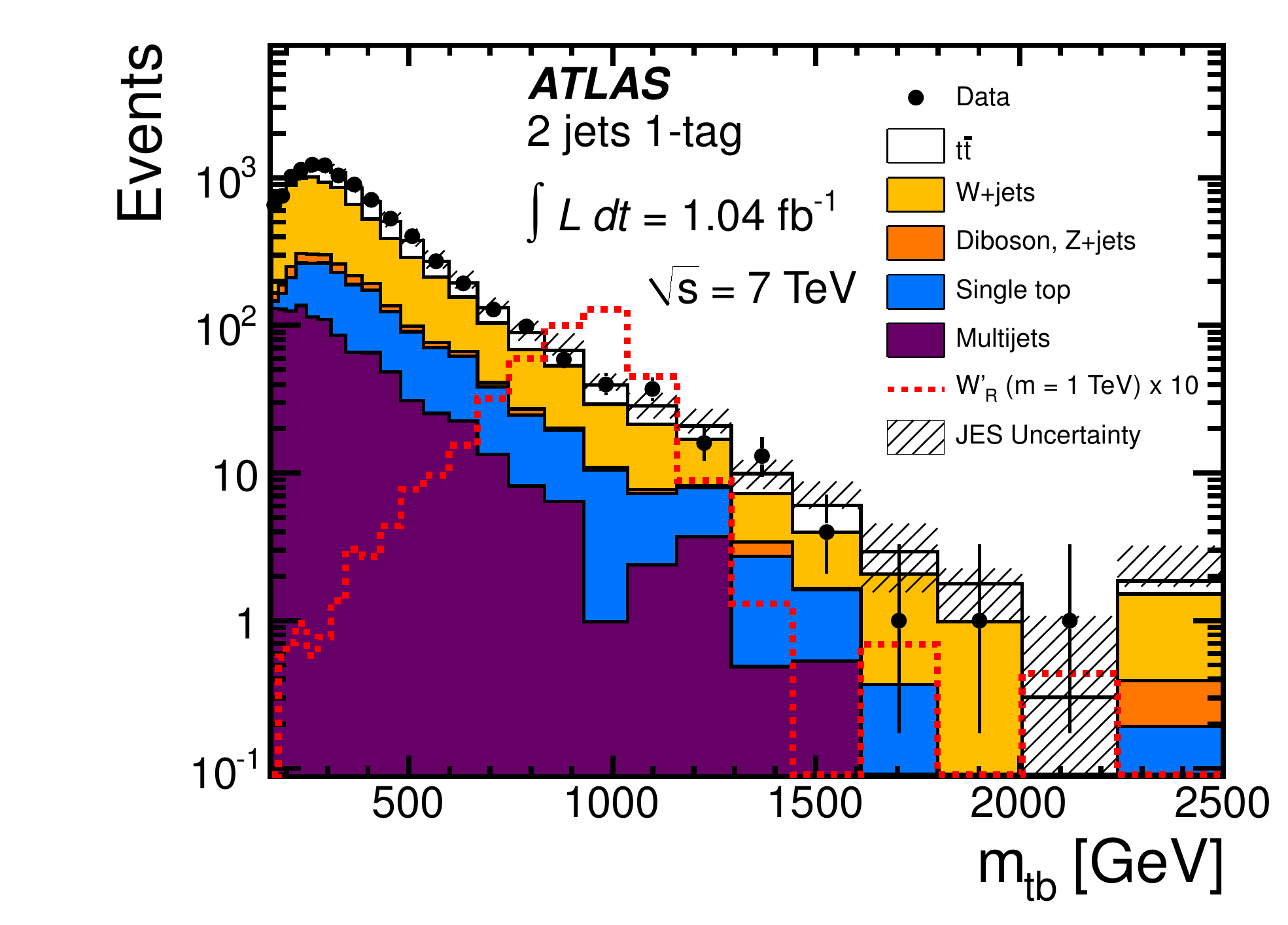,height=2.1in}
\epsfig{file=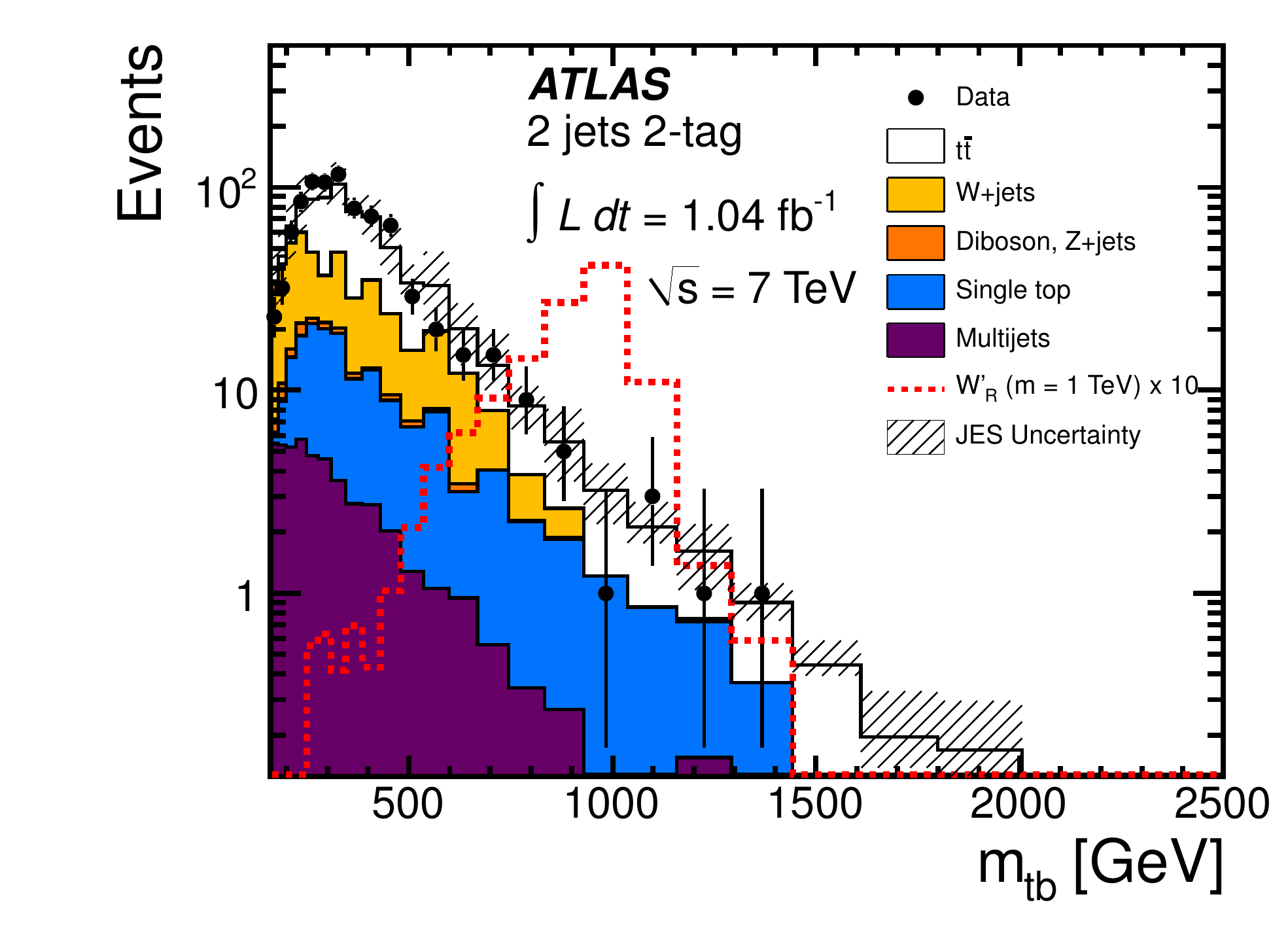,height=2.1in}
a)\hspace*{0.5\textwidth} b)
\caption{\label{fig:invariant}
The invariant mass distribution of $tb$ system, where data is compared to Standard Model expectations for a) single-tagged events b) two-jet events~\cite{Aad:2012ej}.}
\end{center}
\end{figure}

The search for the $tb$ resonance events was performed using the invariant mass distribution of the $tb$ quarks system after applying all cuts, then data and MC distributions were compared for the two channels.
Figure~\ref{fig:invariant}a and Figure~\ref{fig:invariant}b show the $tb$ invariant mass distributions where one can see that data agrees very well with the background modelling. 
Using a Bayesian statistical method, the cross section upper limits at 95\% C.L. were calculated for $W^{'}$ boson decaying into $tb$ quarks for different masses of the $W^{'}$ boson.  This was  done using the binned likelihood of the $tb$ invariant mass distribution of the two channels. The observed cross section upper limits varied between 6.1~pb to 1.0~pb for the $W^{'}$ boson masses between 0.5~TeV and 2~TeV, which were used to calculate the mass lower limit of the $W^{'}$ boson at 95\% C.L. to be 1.13~TeV. 

CMS Collaboration searched for the $W^{'}$ boson by looking at the $tb$ resonance using 5 ${\rm fb}^{-1}$ of data collected in 2011 at $\sqrt{s}=7$~TeV~\cite{:2012sc}. The search for $W^{'}$ boson was performed assuming pure right handed couplings, pure left handed couplings 
and a mixture of the two to the fermions. Since the $W^{'}$ boson with left handed couplings can couple to fermions the same as the SM $W$ boson, there is an interference between the SM  single top quark production 
and the $tb$ coming from the $W^{'}$ decay leading to a contribution from the SM  single top quark of  5-20\%.

The same cuts were applied to select the $W'$ boson events candidates as ATLAS, except the cuts applied to the lepton selection which were different for muon and electron channels. 
Unlike ATLAS, CMS selected events with at least two high $p_{\rm T}$ jets where one jet was identified as a jet originated from a $b$ quark.  
To further suppress the background, the $p_{\rm T}$ of the reconstructed top quark was required to be greater than 75~GeV,  the $p_{\rm T}$ of first leading jet and sub-leading jet to be greater than 100~GeV and  
the reconstructed top quark mass between 130~GeV and 210~GeV.

The signal extraction was done in two ways, by using the invariant mass distributions of the $tb$ system, and by using the boosted decision tree (BDT) distribution which was built from many input variables as was done for the FCNC neural network analysis.  
In the invariant mass analysis, the top quark system was first built from the $W$ boson and a jets giving the closest mass to the known top quark mass, then the $W^{'}$ was constructed from the top quark and the highest $p_{\rm T}$ jet from the remaining jets.  In the BDT analysis, 39 input variables were used to train the BDT for each mass point of each type of coupling $W^{'}_{R}$, $W^{'}_{L}$ and $W^{'}_{LR}$ and for both electron and muon channels.

No evidence for $W^{'}$ boson events was found, and upper limits at 95\% C.L. on the production cross sections was set using the CLs statistical method. 
The binned likelihood was calculated using the full $tb$ invariant mass distributions and BDT distributions in the invariant mass and BDT analyses, respectively. 
The rate normalisation uncertainties and shape distortions of the templates were considered and included in the likelihood assuming  log-normal priors.
 
It found that BDT analyses predicted more sensitive results. The cross section upper limits at 95\% C.L were calculated for each mass value and used to set the mass lower limit at 95\% C.L. to be 1.85~TeV 
for both $W^{'}_{R}$ and $W^{'}_{L}$ ignoring the interference with SM. For the first time, limits on the cross sections were converted to limits on the coupling strengths assuming arbitrary combinations of the left and right handed couplings. This was done using invariant mass analysis results assuming that the $W^{'}$ couples with the same coupling strengths to all generations.

\begin{figure}[htp]
\begin{center}
\epsfig{file=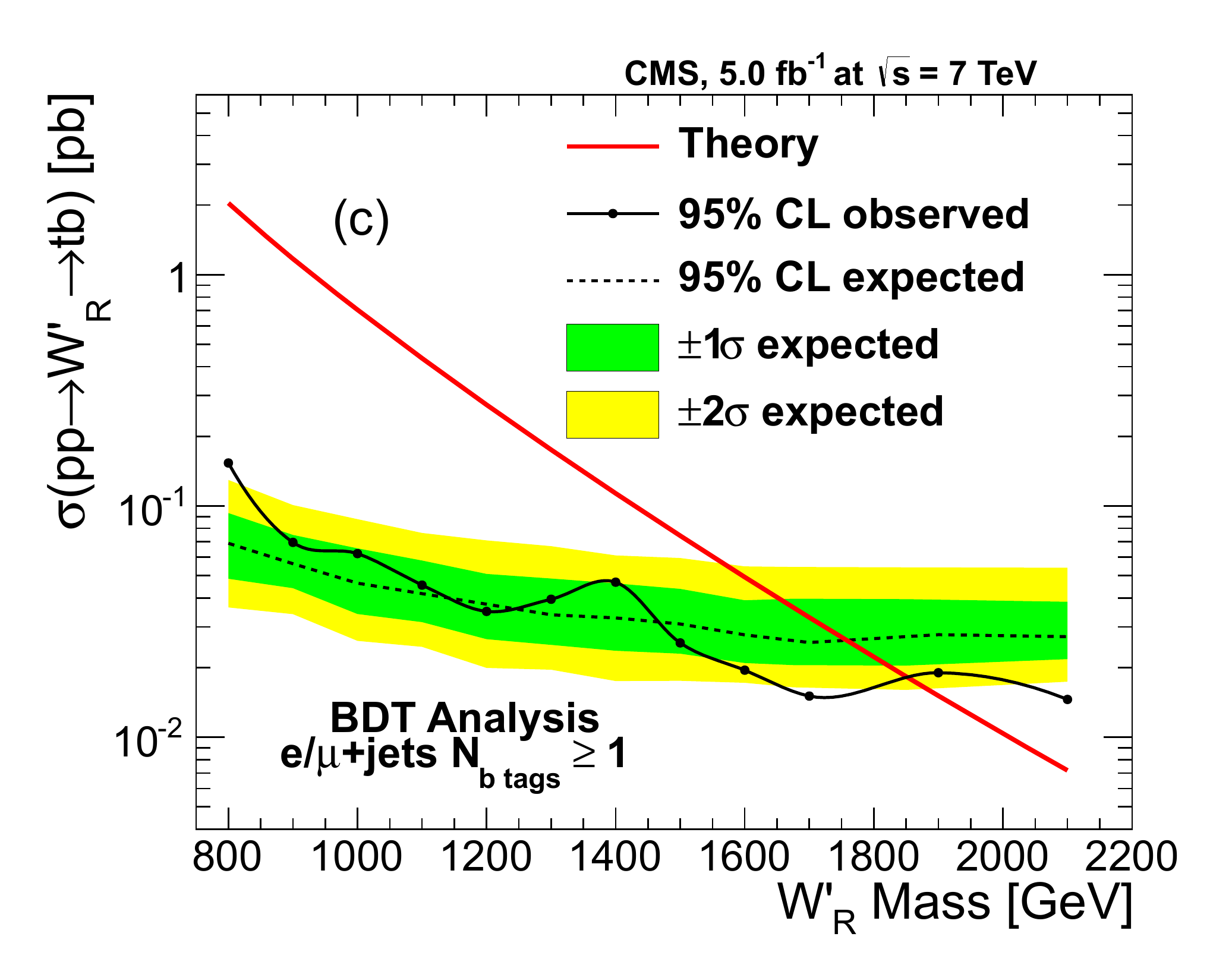,height=2.1in}
\epsfig{file=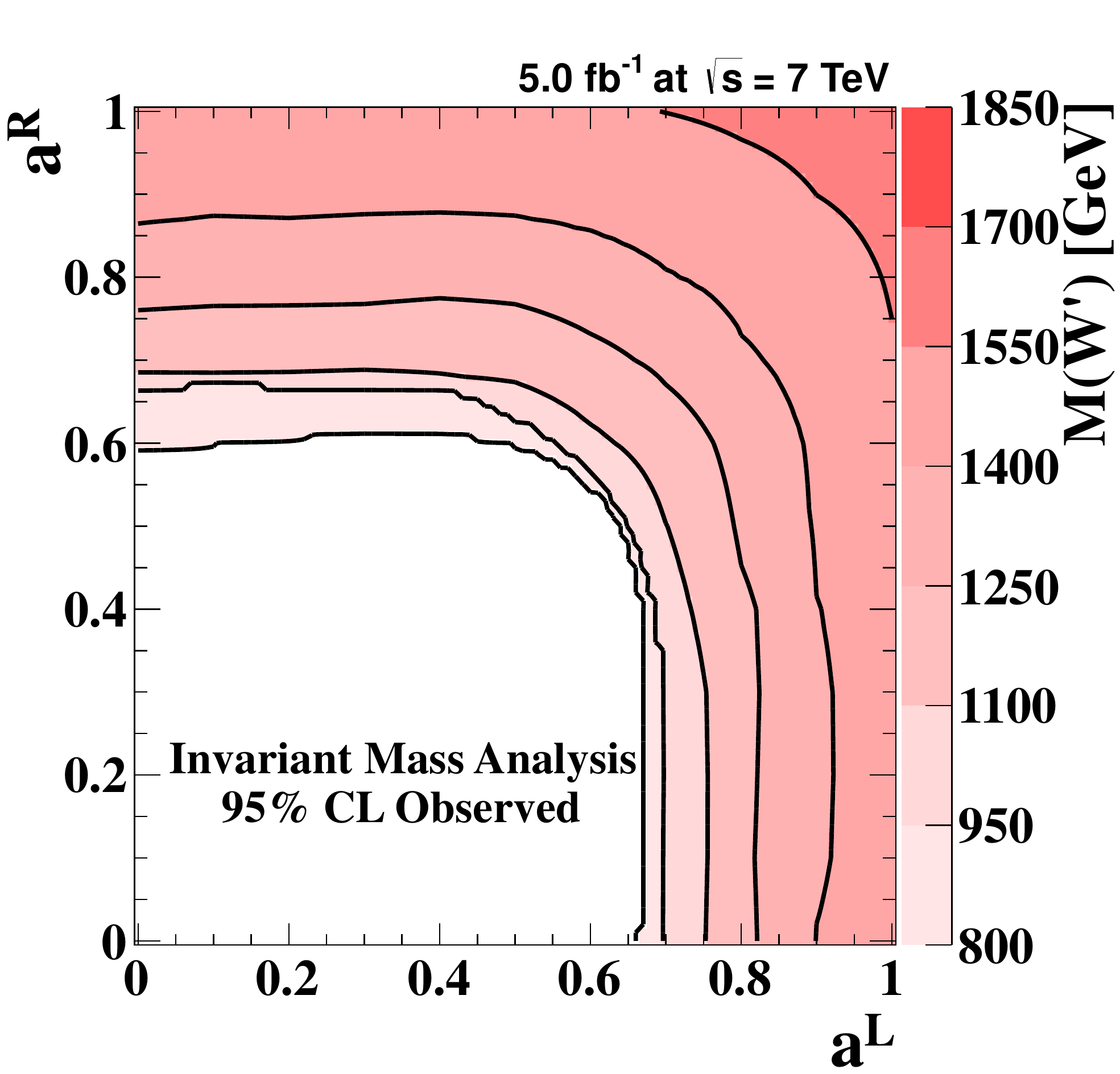,height=2.1in}
a)\hspace*{0.35\textwidth} b)
\caption{\label{fig:CMSresults}a) The measured upper limits at 95\% C.L. on the production cross sections times branching fractions of right handed $W^{'}$ bosons obtained using the BDT for muon and 
electron channels combined. b) The upper limits at 95\% C.L. on the left handed $a^L$ and right handed $a^R$ couplings assuming an arbitrary combination converted from the cross section upper limits calculated for each 
mass point of  the $W^{'}$ boson. This was performed using the results of the invariant mass analysis for muon and electron channels combined~\cite{:2012sc}.}
\end{center}
\end{figure}


\def\Discussion{
\setlength{\parskip}{0.3cm}\setlength{\parindent}{0.0cm}
     \bigskip\bigskip      {\Large {\bf Discussion}} \bigskip}
\def\speaker#1{{\bf #1:}\ }
\def\endDiscussion{}

\end{document}

%% file: econfmacros.tex
\def\beq{\begin{equation}}
\def\eeq#1{\label{#1}\end{equation}}
\def\eeqn{\end{equation}}

\def\beqa{\begin{eqnarray}}
\def\eeqa#1{\label{#1}\end{eqnarray}}
\def\eeqan{\end{eqnarray}}

\let\bar=\overbar

\def\Dslash{\not{\hbox{\kern-4pt $D$}}}
\def\dslash{\not{\hbox{\kern-2pt $\del$}}}

\def\msb{{\bar{\ssstyle M \kern -1pt S}}}

